\documentclass[fleqn,10pt]{wlscirep}

\usepackage{xr}
\usepackage{multirow}

\title{How morphological development can guide evolution}

\author[1,*]{Sam Kriegman}
\author[1]{Nick Cheney}
\author[1]{Josh Bongard}
\affil[1]{University of Vermont, Department of Computer Science, Burlington, VT, USA}

\affil[*]{sam.kriegman@uvm.edu}

\begin{abstract}

Organisms result from adaptive processes interacting across different time scales. 
One such interaction is that between development and evolution. 
Models have shown that development sweeps over several traits in a single agent, sometimes exposing promising static traits. 
Subsequent evolution can then canalize these rare traits. 
Thus, development can, under the right conditions, increase evolvability. 
Here, we report on a previously unknown phenomenon when embodied agents are allowed to develop and evolve: Evolution discovers body plans robust to control changes, these body plans become genetically assimilated, yet controllers for these agents are not assimilated. 
This allows evolution to continue climbing fitness gradients by tinkering with the developmental programs for controllers within these permissive body plans. 
This exposes a previously unknown detail about the Baldwin effect: instead of all useful traits becoming genetically assimilated, only traits that render the agent robust to changes in other traits become assimilated. 
We refer to this as \textit{differential canalization}.
This finding also has implications for the evolutionary design of artificial and embodied agents such as robots: robots robust to internal changes in their controllers may also be robust to external changes in their environment, such as transferal from simulation to reality or deployment in novel environments.

\end{abstract}

\begin{document}

\flushbottom
\maketitle

\thispagestyle{empty}

\section*{Introduction}
\label{sec:intro}

The shape of life changes on many different time scales.
From generation to generation, populations gradually increase in complexity and relative competency.
At the individual level, organisms grow from a single-celled egg and exhibit extreme postnatal change as they interact with the outside world during their lifetimes.
At a faster time scale still, organisms behave such as to survive and reproduce.

Many organisms manifest different traits as they interact with their environment.
It seems wasteful not to utilize this extra exploration to speed the evolutionary search for good genotypes. 
However, to communicate information from these useful but temporary traits to the genotype requires inverting the generally very complex, nonlinear and stochastic mapping from DNA to phenotype.
Inverting such a function would be exceedingly difficult to compute.
Organisms can, however,
pass on their particular capacity to acquire certain characteristics. 
Thus phenotypic plasticity can affect the direction and rate of evolutionary change by influencing selection pressures.
Although this phenomenon was originally described by Baldwin \cite{baldwin1896new}, Morgan \cite{morgan1896modification} and Waddington \cite{waddington1942canalization}, among others, it has become known as `the Baldwin effect'.
In Baldwin's words:
`the most plastic individuals will be preserved to do the advantageous things for which their variations show them to be the most fit, and the next generation will show an emphasis of just this direction in its variations' \cite{baldwin1896new}.
In a fixed environment, when the `advantageous thing' to do is to stay the same, selection can favor genetic variations which more easily, reliably, or quickly produce these traits. 
This can lead to the genetic determination of a character which in previous generations needed to be developed or learned.

Thirty years ago, Hinton and Nowlan \cite{hinton1987learning} provided a simple computational model of the Baldwin effect that clearly demonstrated how phenotypic plasticity could, under certain conditions, speed evolutionary search without communication to the genotype.
They considered the evolution of a bitstring that is only of value when perfectly matching a predefined target string.
The search space therefore has a single spike of high fitness with no slope leading to the summit.
In such a space, evolution is no better than random search.

Hinton and Nowlan then allowed part of the string to randomly change at an additional (and faster) developmental time scale.
When the genetically specified (nonplastic) portion of the string is correct, there is a chance of discovering the remaining portion in development.	
The speed at which such individuals tend to find the good string will be proportional to the number of genetically determined bits.
When the target string is found, development stops and the individual is rewarded for the amount of remaining developmental time.
This has the effect of creating a gradient of increasing fitness surrounding the correct specification that natural selection can easily climb by incrementally assimilating more correct bits to the genotype.

Hinton and Nowlan imagined the bitstring as specifying the connections of a neural network in a very harsh environment.
We are also interested in this interaction of subsystems unfolding at different time scales, but consider an embodied agent situated in a physically-realistic environment rather than an abstract control system.
This distinction is important as it grounds our hypotheses in the constraints and opportunities afforded by the physical world.
It also allows us to investigate how changes in morphology and control might differentially affect the direction or rate of evolutionary search.
More specifically, it exposes the previously unknown phenomenon of differential canalization reported here.

Inspired by Hinton and Nowlan, Floreano and Mondada \cite{floreano1996evolution} explored the interaction between learning and evolution in mobile robots with a fixed body plan but plastic neural control structure.
They noted that the acquisition of stable behavior in ontogeny did not correspond to stability (no further change) of individual synapses, but rather was regulated by continuously changing synapses which were dynamically stable.
In other words, agents exploited this ontogenetic change for behavior, and this prevented its canalization.
In this paper, we structure development in a way that restricts its exploitation for behavior and thus promotes the canalization of high performing static phenotypes.
Also, the robot's body plan was fixed in Floreano and Mondada's experiments\cite{floreano1996evolution}, whereas in the work reported here, evolution and development may modify body plans.

Several models that specifically address morphological development of embodied agents have been reported in the literature \cite{
dellaert1996developmental,
Eggenberger97,
Bongard01,
miller2004evolving,
doursat2009organically
}.
However, the relationship between morphological development and evolvability is seldom investigated in such models.
Moreover, there are exceedingly few cases that considered postnatal change to the body plan of an agent (its resting structural form) as it behaves and interacts with the environment through physiological functioning (at a faster time scale).

We are only aware of four cases reported in the literature in which a simulated robot's body was allowed to change while it was behaving.
In the first two cases \cite{ventrella1998designing, komosinski2003framsticks}, it was not clear whether this ontogenetic morphological change facilitated the evolution of behavior.
Later, Bongard \cite{bongard2011morphological} demonstrated how such change could lead to a form of self-scaffolding that smoothed the fitness landscape and thus increased evolvability.
This ontogenetic change also exposed evolution to a wider range of sensor-motor contingencies, which increased robustness to novel environments.
More recently, Kriegman \textit{et al.} \cite{kriegman2017minimal} showed how development can sweep over a series of body plans in a single agent, and subsequent heterochronic mutations canalize the most promising body plan in more morphologically-static descendants.

We are not aware of any cases reported in the literature to date in which a simulated robot's body and control are simultaneously allowed to change while it is behaving.
In this paper, we investigate such change in the morphologies and controllers of soft robots as they are evolved for coordinated action in a simulated 3D environment.
By morphology we mean the current state of a robot’s shape, which is slowly changed over the course of its lifetime by a developmental process.
We distinguish this from the controller, which sends propagating waves of actuation throughout the individual, which also affects the instantaneous shape of the robot but to a much smaller degree.
We here refer to these two processes as `morphology' and `control'.
As both processes change the shape, and thus behavior, of the robot, this distinction is somewhat arbitrary.
However, the central claim of this paper, which is that some traits become canalized while others do not, is not reliant on this distinction.

We use soft robots because they provide many more degrees of morphological freedom compared to traditional robots composed of rigid links connected by rotary or linear actuators.
This flexibility allows soft robots to accomplish tasks that would be otherwise impossible for their rigid-bodied counterparts, such as squeezing through small apertures \cite{Cheney:2015:ESR:2739480.2754662} or continuously morphing to meet different tasks.
Recent advancements in materials science are enabling the fabrication of 3D-printed muscles \cite{miriyev2017soft} and nervous systems \cite{wehner2016integrated}.
However, there are several challenges to the field of soft robotics, including an overall lack of design intuition:
What should a robot with nearly unbounded morphological possibility look like, and how can it be controlled?
Controllability often depends on precision actuation and feedback authority, but these properties are difficult to maintain in soft materials in which motion in one part of the robot can propagate in unanticipated ways throughout its body \cite{lipson2014challenges}.

We present here a minimally complex but embodied model of morphological and neurological development.
This new model represents an alternative approach to the challenging problem of soft robot design and presents an \textit{in silico} testbed for hypotheses about evolving and developing embodied systems.
This model led to the discovery of differential canalization and how it can increase evolvability.

\externaldocument{methods}
\externaldocument{figures}

\begin{figure}[t]
\centering
\includegraphics[width=0.8\linewidth]{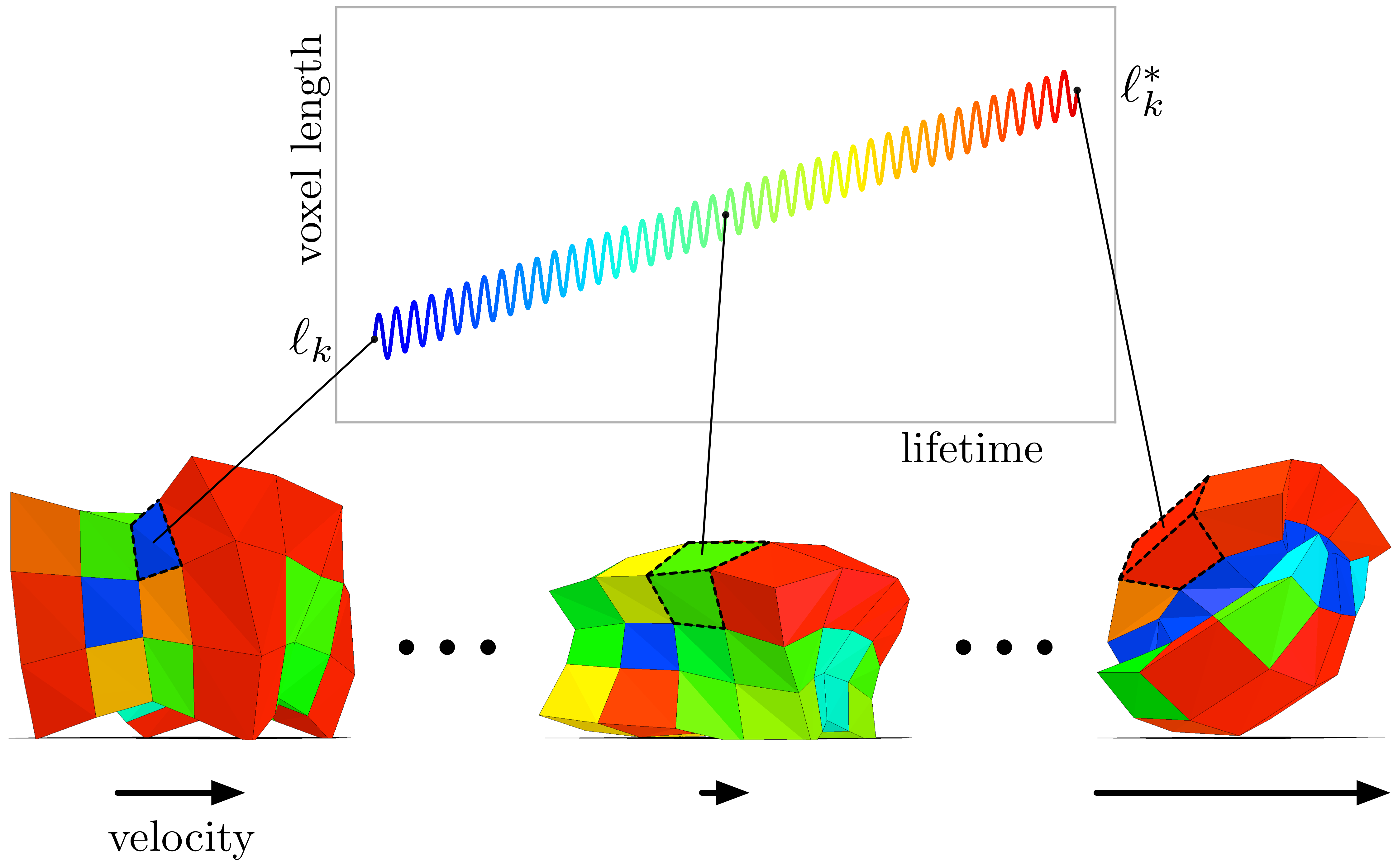}
\caption{\label{fig:blueprint}\textbf{Modeling development.} An evolved soft robot changes its shape during its lifetime (postnatal development), from a walking quadruped into a rolling form. 
Evolution dictates how a robot's morphology develops by setting each voxel's initial ($\ell_k$) and final ($\ell^*_k$) resting length.
The length of a single voxel $k$ is plotted to illustrate its (slower) growth and (faster) actuation processes.
Voxel color indicates the current length of that cell: the smallest voxels are blue, medium sized voxels are green, and the largest voxels are red.
As robots develop and interact with a physically realistic environment, they generate heterogeneous behavior in terms of instantaneous velocity (bottom arrows).
Soft robot evolution, development and physiological functioning can be seen in Supplementary \href{https://youtu.be/Ee2sU-AZWC4}{Video S1}.
}
\end{figure}

\begin{figure}[t]
\includegraphics[width=\linewidth]{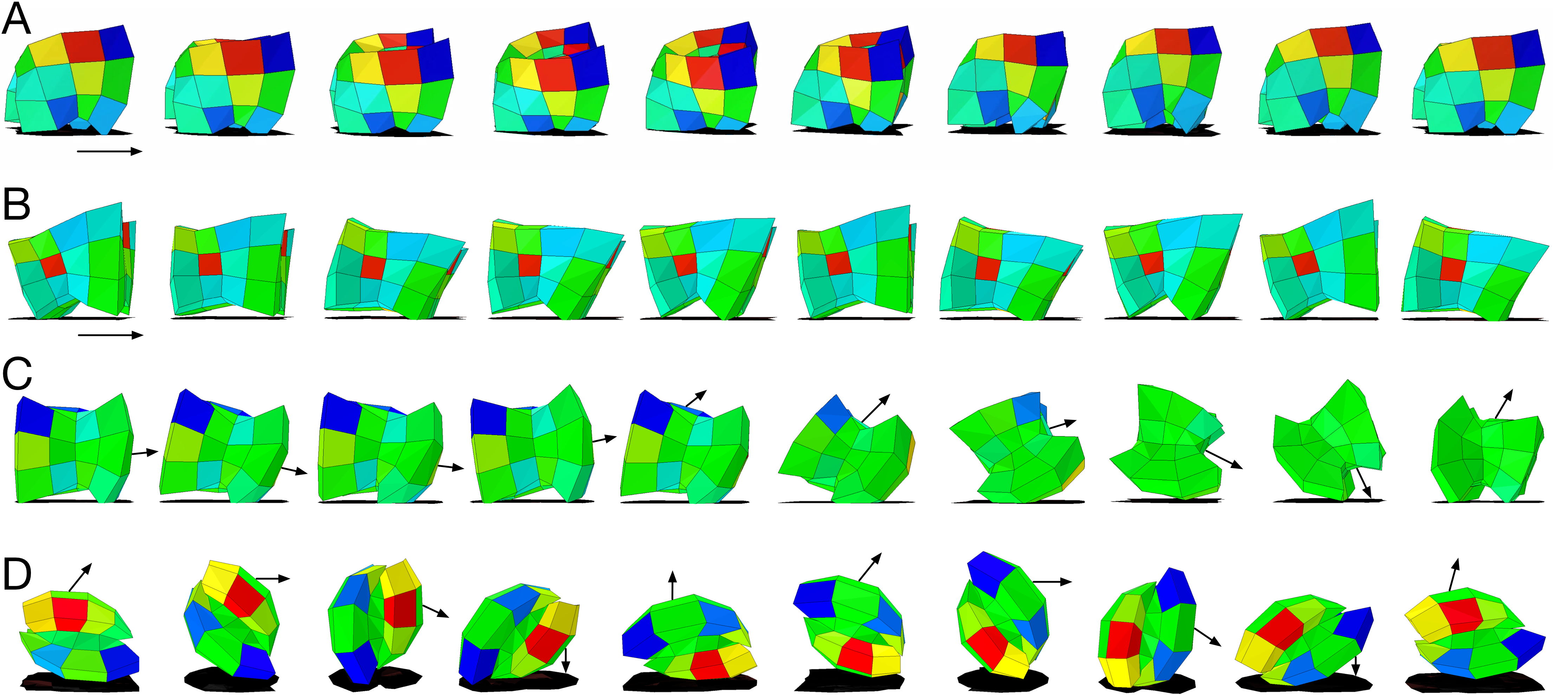}
\caption{\label{fig:trot-gallop-roll}\textbf{Evolved behavior.}
Each row depicts a different evolved robot moving from left to right. 
Voxels in this figure are colored by the amount of subsequent morphological development remaining at that cell: blue indicates shrinking voxels $(\ell_k > \ell_k^*)$, red indicates growing voxels $(\ell_k < \ell_k^*)$, green indicates little to no change either way $(\ell_k \cong \ell_k^*)$.
(A) An evolved trotting soft quadruped with a two-beat gait synchronizing diagonal pairs of legs. 
(B) A galloping adult robot which goes fully airborne mid-gait.
(C) A galloping juvenile robot which develops into a rolling adult form. 
(D) A rolling juvenile robot at 10 points in ontogeny immediately after birth.
Arrows indicate the general directionality of movement,  
but this is more precisely captured by
Supplementary \href{https://youtu.be/Ee2sU-AZWC4}{Video S1}. 
}
\end{figure}

\begin{figure}[t]
\centering
\includegraphics[width=0.9\linewidth]{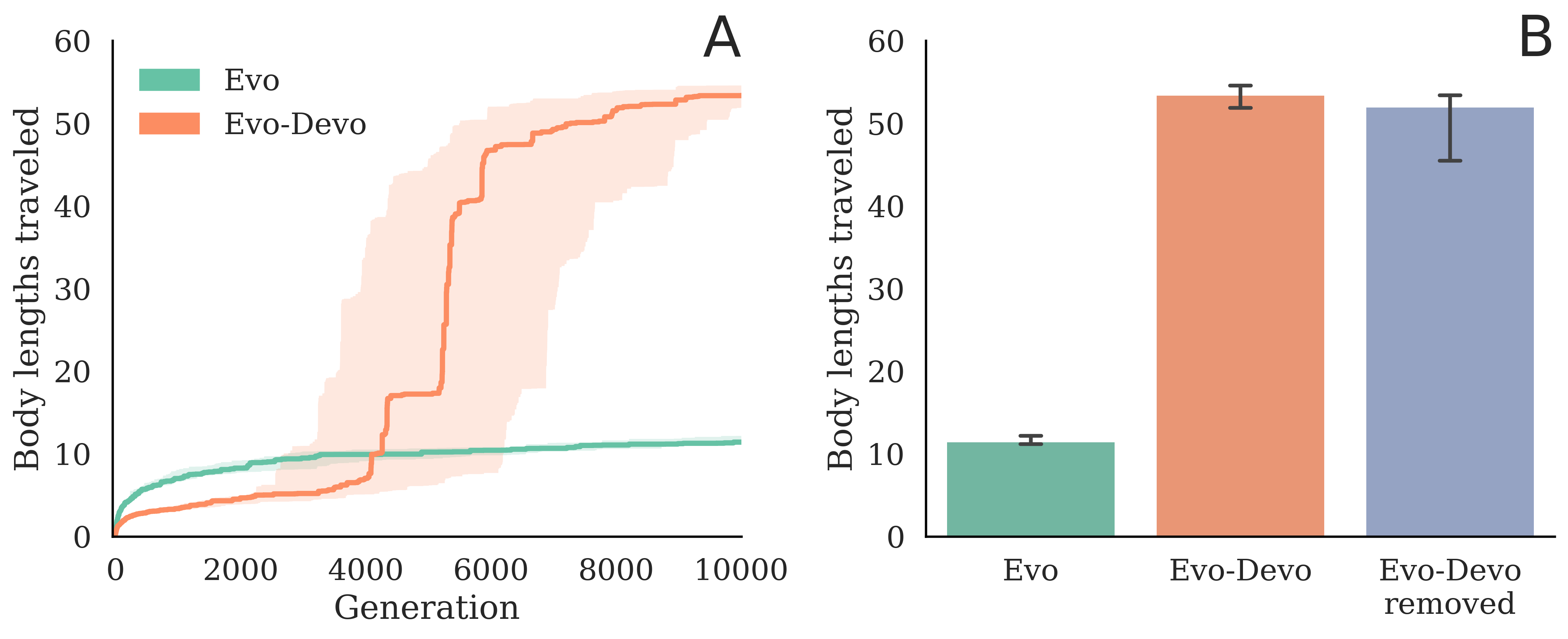}
\caption{\label{fig-fitness}\textbf{Evolvability and development.} Morphological development drastically increases evolvability (A), even when development is manually removed from the evolved systems (the run champions) by setting the final parameter values equal to their starting values ($\ell_k^*=\ell_k$ and $\phi_k^*=\phi_k$), in each voxel (B).
Median fitness is plotted with 95\% bootstrapped confidence intervals for three treatments: evolving but non-developmental robots (Evo), evolving and developing robots (Evo-Devo), and evolving and developing robots evaluated at the end of evolution with their development removed (Evo-Devo removed).
Fitness of just the final, evolved populations (at generation 10000) are plotted in B.}
\end{figure}

\begin{figure}[t]
\centering
\includegraphics[width=\linewidth]{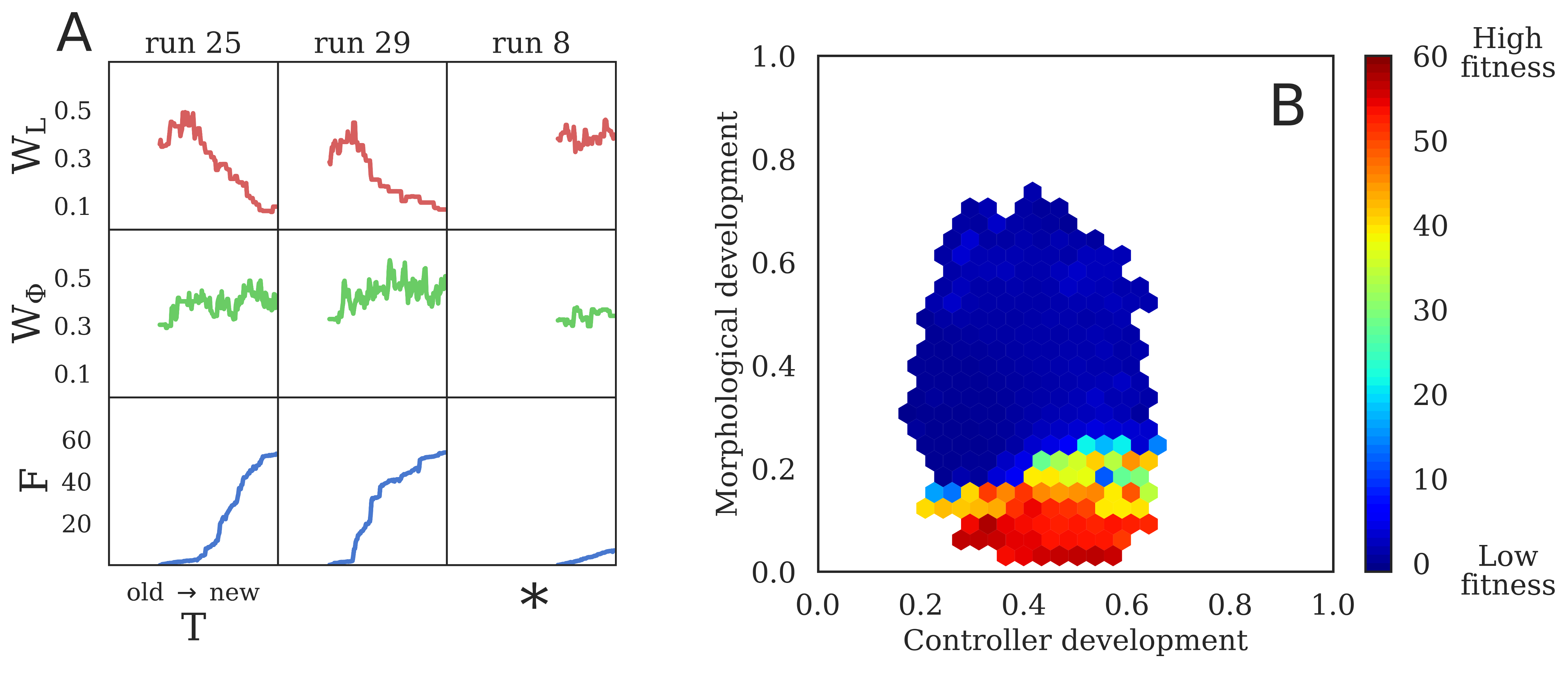}
\caption{\label{fig-correlation}\textbf{Differential canalization.} 
Developmental windows (i.e.~the total lifetime developmental change) for morphology, $W_L$ (see Equation \ref{eq-WL}), and controller, $W_{\Phi}$ (see Equation \ref{eq-WPhi}), alongside fitness $F$.
(A) Three representative lineages taken from Supplementary Fig. S1, which displays the lineages of all 30 Evo-Devo run champions. Evolutionary time $T$ moves from the oldest ancestor (left) to the run champion (right). A general trend emerges wherein lineages initially increase their morphological development in $T$ (rising red curves) and subsequently decrease morphological development to almost zero (falling red curves). Five of the 30 evolutionary trials, annotated by {\Large $\ast$}, fell into a local optima.
(B) Median fitness as a function of morphology and controller development windows $(W_L,\; W_{\Phi})$, for all Evo-Devo designs evaluated. 
Overall, the fastest designs tend to have small amounts of morphological development, but are free to explore alternative control policies.}
\end{figure}

\begin{figure}[t]
\centering
\includegraphics[width=\linewidth]{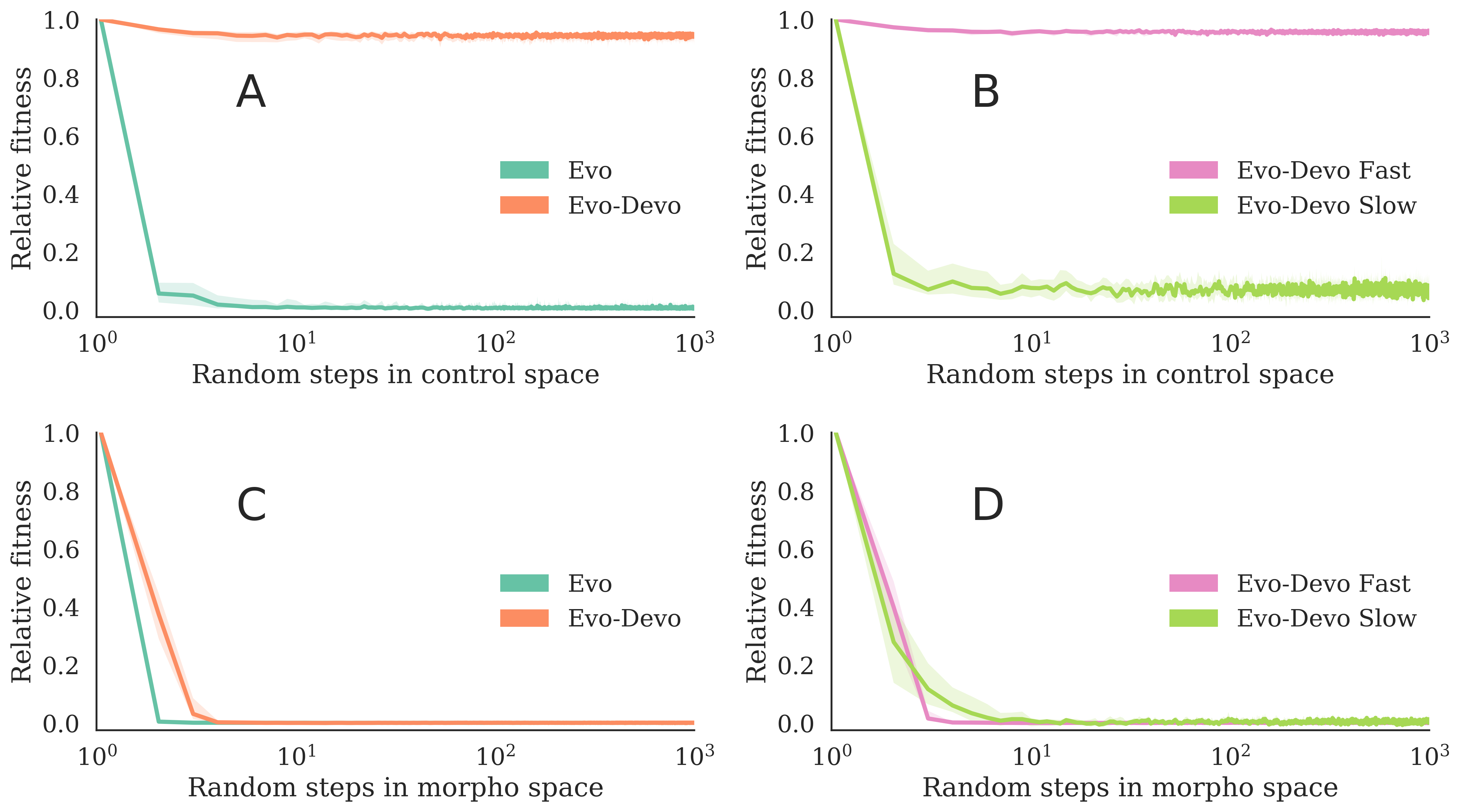}
\caption{\label{fig-random-walks}\textbf{Sensitivity to morphological and control mutations.}
Ten random walks were taken from each run champion.
(A) Successive \textit{control} mutations to the Evo and Evo-Devo run champions.
(B) The previous Evo-Devo results separately for fast and slow design types.
(C) Successive \textit{morphological} mutations to the Evo and Evo-Devo run champions.
(D) The previous Evo-Devo results separately for fast and slow design types. Medians plotted with 99\% confidence intervals.
The faster Evo-Devo robots tend to possess body plans that are robust to control mutations.}
\end{figure}

\begin{figure}[t]
\centering
\includegraphics[width=0.9\linewidth]{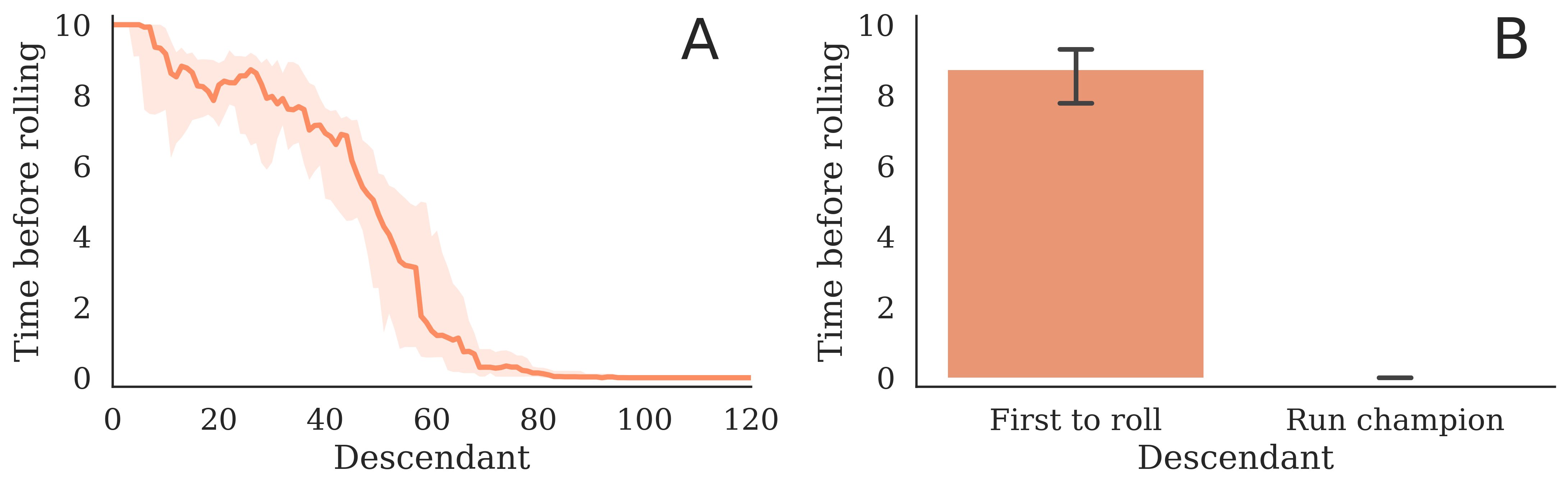}
\caption{\label{fig-discovery}\textbf{Late onset discoveries.} Ontogenetic time before the discovery of rolling over, taken from the lineages of the best robot from each of the 25 Evo-Devo trials that produced a rolling design. Median time to discovery, with 95\% C.I.s, for (A) the lineage from the most distant ancestor $(T=0)$ to more recent descendants, and (B) the first ancestor to roll over compared to the final run champion. 
Rolling over is measured from the first time step the top of the robot touches the ground, rather than after completely rolling over. 
The first ancestors to roll over tend to do so at the end of their lives, their descendants tend to roll sooner in life, and the final run champions all begin rolling immediately at birth. 
These results are a consequence of dependent time steps: 
because mutational changes affect all downstream steps, their phenotypic impact is amplified in all but the terminal stages of development. Thus, late onset changes can provide exploration in the search space without breaking rest-of-life functionality, and subsequent evolution can gradually assimilate this trait to the start of development.}
\end{figure}

\section*{Results}
\label{sec:results}

We consider a locomotion task, over flat terrain, for soft robots composed of a $4\times4\times3$ grid of voxels (Fig. \ref{fig:blueprint}).
Robots are evaluated for 40 actuation cycles at 4 Hz, yielding a lifetime of ten seconds. Fitness is taken to be distance traveled measured in undeformed body lengths (four unit voxels, i.e.~4 cm). 
Example robots are shown in Figures \ref{fig:blueprint} and \ref{fig:trot-gallop-roll}, and Supplementary \href{https://youtu.be/Ee2sU-AZWC4}{Video S1}.

All experiments were performed in the open-source soft-body physics simulator  
\textit{Voxelyze} \cite{hiller2014dynamic}. 
Voxelyze simulates soft materials using two elements: particles and beams. 
A particle is a point mass with rotational inertia.
A spring-like beam (with translational and rotational stiffness) connects two adjacent particles.
Each particle is connected to at most six neighbors (above, below, front, back, left, and right), on a cartesian grid.
Local material properties are stored at particles and averaged across shared beams.
Finally, a voxel mesh is drawn around each particle for visualization, such that adjacent voxels touch at the center of their shared beam.
More details about how this is actually implemented are given by Hiller and Lipson \cite{hiller2014dynamic}.

The morphology of a robot is given by the resting (beam) length stored at each voxel (Fig. \ref{fig:blueprint}).
However the shape and volume of each voxel is changed by external forces from the environment and internal forces via behavior. 
The morphology of a robot is denoted by the $4\times4\times3=48$-element vector $\ell$, where each element is the resting length stored at that voxel (with possible values within $1.0 \pm 0.75\; \text{cm}$).
Like most animals, our robots are bilaterally symmetrical.
We built this constraint into our robots because bilateral symmetry is known to help with forward locomotion
\cite{grabowsky1994symmetry}.
The lefthand $2\times4\times3=24$ resting voxel lengths are reflected on the other, righthand side of the midsagittal line, yielding 24 independent resting lengths.

The controller, however, is not constrained to be symmetrical since many behaviors, even for symmetric morphologies, consist of asymmetric gaits, and is given by the phase offset of each voxel from a global oscillating signal with an amplitude of 0.14 cm. 
The controller is denoted by the 48-element vector $\phi$, where each element is the phase offset of that voxel (with possible values within $0 \pm \pi/2$).

We investigated the impact of development in this model by comparing two experimental variants: Evo and Evo-Devo (schematized in Supplementary Fig. \hyperref[fig:S2]{S2}). 
The control treatment, \textbf{Evo}, lacks development and therefore maintains a fixed morphology and control policy in a robot as it behaves over its lifetime. 
Two parameters per voxel are sufficient to specify an evolved robot at any time $t$ in its lifetime: its morphology $\ell_k$, and controller $\phi_k$.
An evolutionary algorithm optimizes 24 morphological and 48 control parameters.

The experimental treatment \textbf{Evo-Devo} evolves a developmental program rather than a static phenotype (Fig. \ref{fig:blueprint}). 
For each parameter in an Evo robot, an Evo-Devo robot has two: its starting and final value.
The evolutionary algorithm associated with the Evo-Devo treatment thus optimizes 48 morphological and 96 control parameters.
The morphology and controller of the $k$-th voxel change linearly from starting to final values, throughout the lifetime of a developing robot.
The endpoint parameters are denoted by asterisks: the controller develops from $\phi$ to $\phi^*$, the morphology develops from $\ell$ to $\ell^*$.
The starting and final points of development are predetermined by a genome which in turn fixes the direction (compression or expansion) and rate of change for each voxel.
Development is thus \textit{ballistic} in nature rather than adaptive, as it cannot be influenced by the environment (Equation \ref{eq:ballistic-devo}).

For both treatments we conducted 30 independent evolutionary trials.
At the end of evolutionary optimization, the non-developmental robots (Evo) tend to move on average with a speed of 10 body lengths in 10 seconds, or 1 length/sec. The evolved and developing robots (Evo-Devo) tend to move at over 5 lengths/sec (Fig. \ref{fig-fitness}A).
To ensure evolved and developing robots are not exploiting some unfair advantage conferred by changing body plans and control policies unavailable to non-developmental robots, we manually remove their development by setting $\ell^*=\ell$ and $\phi^*=\phi$, which fixes the structure of their morphologies and controllers at birth ($t=0$) (Equation \ref{eq:ballistic-devo}). 
That is to say, we convert the evolved Evo-Devo robots into Evo robots (Equation \ref{eq:ballistic-devo} reduces to Equation \ref{eq:no-devo}).
The resulting reduced robots suffer only a slight (and statistically non-significant) decrease in median speed and still tend to be almost five times faster than the systems evolved without development (Fig. \ref{fig-fitness}B, treatment `Evo-Devo removed').
Ballistic development is therefore beneficial for search but does not provide a behavioral advantage in this task environment.

To investigate this apparent search advantage, we trace development and fitness across the 30 lineages which produced a `run champion': the robot with highest fitness at the termination of a given evolutionary trial (Supplementary Fig. \hyperref[fig:S1]{S1}). 
We measure the amount of ballistic change in each robot|its `ballistic plasticity'|by a statistic we call the \textit{developmental window}.
The developmental window is defined separately for morphology (Equation \ref{eq-WL}) and control (Equation \ref{eq-WPhi}) as the absolute difference in starting and final values summed across the robot and divided by the total amount of possible development, such that 0 and 1 indicate no and maximal developmental change, respectively.
Evo robots by definition have development windows of zero, as do Evo-Devo robots that have had development manually removed.
An Evo-Devo robot with a small developmental window has thus become \textit{canalized} \cite{waddington1942canalization}.

In terms of fitness, there were two observed basins of attraction in average velocity: a slower design type which either trots or gallops at a speed of less than 1 length/sec (Fig. \ref{fig:trot-gallop-roll}A,B and Fig. \ref{fig-correlation}A{\Large$\ast$}), and a faster design type that rolls at 5-6 lengths/sec (Fig. \ref{fig:trot-gallop-roll}C). 
After ten thousand generations, 25 out of a total of 30 Evo-Devo trials (83.3\%) find the faster design, compared to just 6 out of 30 Evo trials (20\%).

\subsection*{Differential canalization.}

Modular systems are more evolvable than non-modular systems because they allow evolution to improve one subsystem without disrupting others \cite{wagner1996perspective,lipson2007principles}.
Modularity may be a property of the way a system is built, or it may be an evolved property.
The robots evolved here are by definition modular because the genes which affect morphology are independent of those which affect its control.
However the more successful Evo-Devo lineages evolved an additional form of modularity, which we term differential canalization:
Some initially developmentally plastic traits become integrated and canalized, while other traits remain plastic.

In the successful Evo-Devo trials, morphological traits were canalized while control traits were not.
Evidence for this is provided in Supplementary Fig. \hyperref[fig:S1]{S1}, which is summarized by Fig. \ref{fig-correlation}A.
Trajectories of controller development (green curves) do not follow any discernible pattern in phylogenetic time, and appear upon visual inspection to be consistent with a random walk or genetic drift.
The trajectories of morphological development (red curves), however, follow a consistent pattern.
The magnitude of morphological development increases slightly, but significantly $(p<0.001)$, before decreasing all the way to the most recent descendant, which is the most fit robot from that trial (the run champion). 
Run champions tend to have much less morphological development than their most distant ancestor $(p<0.001)$, but there is not a significant difference between champion and ancestral controller windows.
Furthermore, this pattern tends to correlate with high fitness: in trials in which this pattern did not appear (runs 6, 8, 16-18), fitness did not increase appreciably over evolutionary time.

This process within the lineages of the run champions is consistent with a more general correlation found in all designs explored during optimization across all runs: Individuals with the highest fitness values tend to have very small amounts of morphological development, while their control policies are free to develop (Fig. \ref{fig-correlation}B).
However, despite the fact that morphological development tends to be canalized in the most fit individuals, it cannot simply be discarded as the non-developmental systems have by definition small morphological windows, and small controller windows, but also low fitness.

To test the sensitivity of the evolved morphologies to changes in their control policies, we applied a random series of control mutations to the Evo and Evo-Devo run champions from each evolutionary trial.
For each run champion, we perform 1000 subsequent random controller mutations that build upon each other in series (a Brownian trajectory in the space of controllers)|and repeat this process ten times for each run champion, each with a unique random seed.
It was found that optimized Evo-Devo robots tend to possess body plans that are much more robust to control mutations than those of Evo robots (Fig. \ref{fig-random-walks}A).
The first control mutation to optimized Evo robots tends to immediately render them immobile, whereas optimized Evo-Devo robots tend to retain most of their functionality even after 1000 successive random changes to their controllers.
Within Evo-Devo designs, the functionality of the 25 fast designs are minimally affected by changes to their control, whereas the five slow designs also tend to break after the first control mutation (Fig. \ref{fig-random-walks}B). 
Thus it can be concluded that these five robots are non-modular: their non-canalized morphologies evolved a strong dependency on their controllers. 
The Evo robots are similarly non-modular: they are brittle to control mutations.

To test the sensitivity of the evolved controllers to changes in their morphologies, we applied the same procedure described in the previous paragraph but with random morphological mutations rather than control mutations.
It was found that both developmental and non-developmental systems tend to evolve controllers that are very sensitive to morphological mutations (Fig. \ref{fig-random-walks}C).
These findings are consistent with those of Cheney \textit{et al.} \cite{cheney2017scalable}, who also reported that robots were more sensitive to changes in their morphology than in their controllers.
Here, the first few morphological mutations to optimized robots, in both treatments, tend to immediately render them immobile.
Within Evo-Devo design types, neither of which canalized development in their controllers (Supplementary Fig. \hyperref[fig:S1]{S1}), both the fast and slow designs possess controllers sensitive to changes in their morphologies (Fig. \ref{fig-random-walks}D).
Thus it can be concluded that the non-canalized controllers evolved a strong dependency on their morphologies. 
Therefore the only trait to be successfully canalized was also the only trait that rendered the agent robust to changes in other traits.

\subsection*{Heterochrony in morphological development.}

The evolutionary algorithm can rapidly discover an actuation pattern that elicits a very small amount of forward movement in these soft robots regardless of the morphology. 
There is then an incremental path of increasing locomotion speed that natural selection can climb by gradually growing legs to reduce the surface area touching the floor and thus friction, and simultaneously refining controller actuation patterns to better match and exploit the morphology (Fig. \ref{fig:trot-gallop-roll}A,B).

There is, however, a vastly superior design partially hidden from natural selection---a `needle in the haystack', to use Hinton and Nowlan's metaphor \cite{hinton1987learning}.
On flat terrain, rolling can be much faster and more efficient than walking, but finding such a design is difficult because the fitness landscape is deceptive.
Rolling over once is much less likely to occur in a random individual than shuffling forwards slightly. And as a population continues to refine walking morphologies and gaits, lineages containing rocking individuals which are close to rolling over, or roll over just once, do not survive long enough to eventually produce a true rolling descendant. 

Development can alter the search space evolution operates in because individuals sweep over a continuum of phenotypes, with different velocities, rather than single static phenotype that travels at a constant speed (Supplementary Fig. \hyperref[fig:S2]{S2}E,J).
The lineages which ultimately evolved the faster rolling design initially increased their morphological plasticity in phylogenetic time as evidenced by the initial upward trends in the red curves in Supplementary Fig. \hyperref[fig:S1]{S1} (summarized by Fig. \ref{fig-correlation}A) which contain a statistically significant difference between their starting and maximum developmental window sizes $(p<0.001)$.
This exposes evolution to a wider range of body plans and thus increases the chance of randomly rolling at least once at some point during the evaluation period.

The peak of morphological plasticity in  Supplementary Fig. \hyperref[fig:S1]{S1} (summarized by Fig. \ref{fig-correlation}A) generally lines up with the start of an increasing trend in fitness (blue curves) and marks the onset of differential canalization.
Rolling just once allows an individual to move further (1 body length) than some early walking behaviors but they incur the fitness penalty of having fallen over and thus not being able to subsequently walk for the rest of the trial. 
Therefore this tends to happen at the very end of ontogeny (Fig. \ref{fig-discovery}), as individuals evolve to `dive' in the last few time steps of the simulation of their behavior, thus incurring an additional increase of fitness over their parent, which does not exhibit this behavior.
Since more rolling incurs more fitness than less rolling, a form of progenesis occurs as heterochronic mutations move $\ell_k$ closer to $\ell_k^*$, for each voxel.
This gradually earlifies rolling from a late onset behavior to one that arises increasing earlier in ontogeny (Supplementary \href{https://youtu.be/Ee2sU-AZWC4}{Video S1}).
As more individuals in the population discover and earlify this rolling behavior, the competition stiffens until eventually individuals which are not born rolling from the start are not fast enough to compete (Fig. \ref{fig:trot-gallop-roll}C,D).

\subsection*{Generality of results.}

\begin{figure}
\centering
\includegraphics[width=\linewidth]{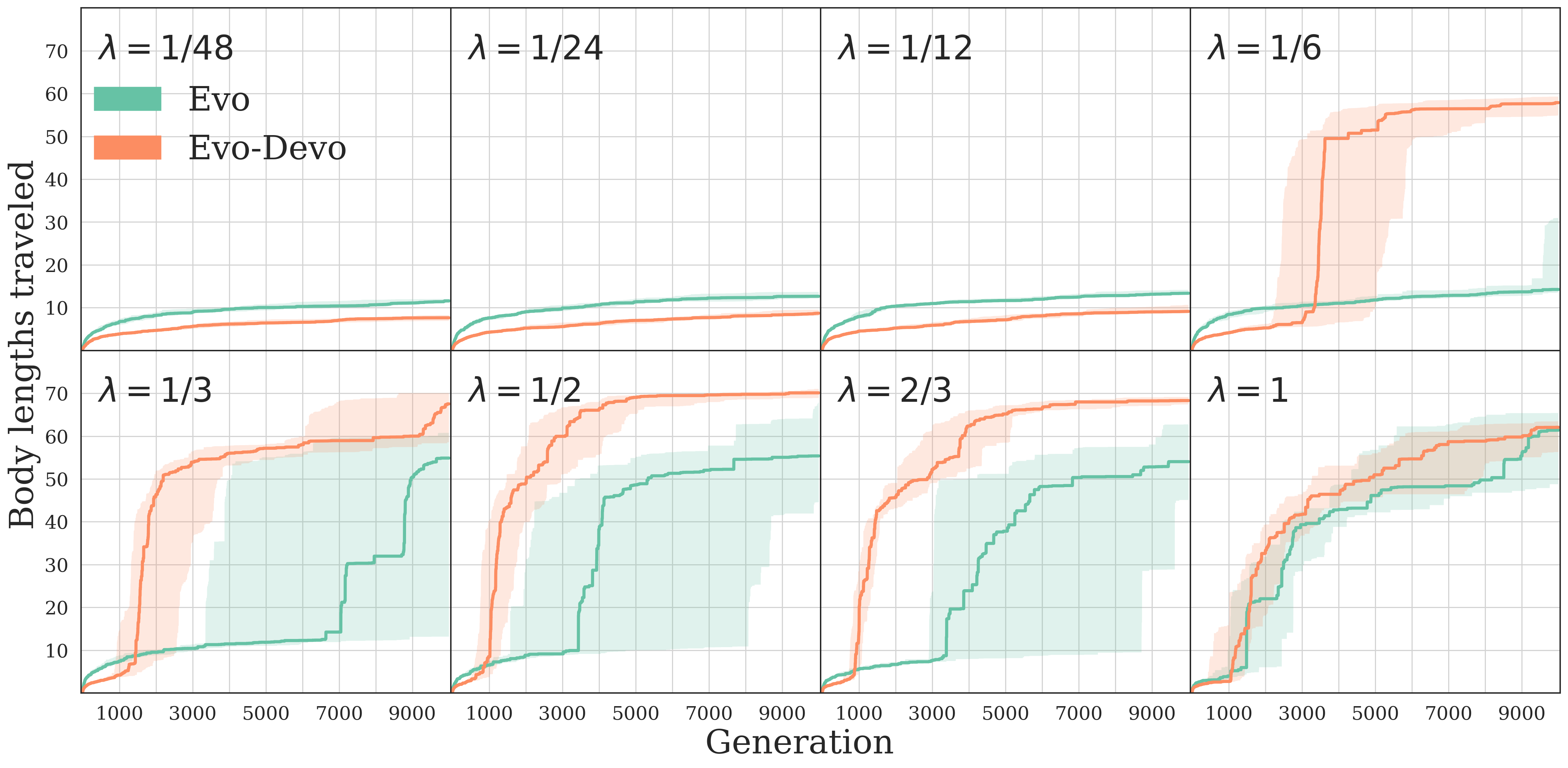}
\caption*{\label{fig:S3}\textbf{{\sf \textbf{Figure S3.}} Mutation rate sweep.} 
Median fitness (with 95\% bootstrapped confidence intervals) under various mutation rates, $\lambda$, a hyperparameter defined in Supplementary Methods which affects the probability a voxel is mutated. 
In the main experiment of this paper, the mutation rate is evolved for each voxel independently, and is constantly changing.
In this mutation rate sweep, $\lambda$ is held uniform across all voxels.
There were two observed basins of attraction in terms of fitness: a slower design that trots/gallops 5-15 body-lengths during the evaluation period, and a faster design type that rolls at 50-70 body-lengths. 
Although higher mutation rates facilitate the discovery of the superior phenotype, once found, lower mutation rates tend to produce more refined and faster robots.
Without development, the search space has a single spike of high fitness. 
One can not do better than random search in such a space.
When $\lambda=1$, optimizing Evo morphologies reduces to random search, and this is the only mutation rate where Evo does not require significantly more generations than Evo-Devo to find the faster design type.
This can be observed for $\lambda\in\{1/6,\, 1/3,\, 1/2,\, 2/3,\, 1\}$, by comparing the generation at which the slopes of the fitness curves increase dramatically.
However, the best two treatments (Evo-Devo at $\lambda=1/2$ and $\lambda=2/3$), as measured by the highest median speed at the end of optimization, have development, and the robots they produced are significantly faster than those produced by random search (Evo with the highest mutation rate)
$(p<0.01)$.
}
\end{figure}

For the results above, as in nature \cite{lynch2010evolution}, the mutation rate of each voxel was left under evolutionary control (self-adaptation).
In an effort to assess the generality of our results, we replicated the experiment described above for various fixed mutation rates (Supplementary Fig. \hyperref[fig:S3]{S3}).
Without development, as in Hinton and Nowlan's case\cite{hinton1987learning}, the search space has a single spike of high fitness. 
One can not do better than random search in such a space.
At the highest mutation rate, optimizing Evo morphologies reduces to random search, and this is the only mutation rate where Evo does not require significantly more generations than Evo-Devo to find the faster design. 
This can be observed in Supplementary Fig. \hyperref[fig:S3]{S3} by comparing the generation at which the slopes of the fitness curves increase dramatically. 
However, the best two treatments, as measured by the highest median speed at the end of optimization, have development, and the robots they produced are significantly faster than those produced by random search (Evo with the highest mutation rate) $(p<0.01)$.

To further test the sensitivity of our results to the various settings of our particular system,
we transcribed the main experiment for a different class of morphologies (rigid bodies) and controllers (neural networks).
Details are provided in Supplementary Fig. \hyperref[fig:S5]{S5} and Supplementary Methods.
The results of this test indicate that differential canalization exists elsewhere, but it does not always increase evolvability.

\section*{Discussion}
\label{sec:discussion}

In these experiments, the intersection of two time scales|slow linear development and rapid oscillatory actuation, as from a central pattern generator|generates positive and negative feedback in terms of instantaneous velocity: the robot speeds up and slows down during various points in its lifetime (Supplementary Fig. \hyperref[fig:S2]{S2}J).
Prior to canalization, unless all of the phenotypes swept over by an individual in development keep the robot motionless, there will be intervals of relatively superior and inferior performance.
Evolution can thus improve overall fitness in a descendant by lengthening the time intervals containing superior phenotypes and reducing the intervals of inferior phenotypes. However, this is only possible if such mutations exist.

We have found here that such mutations do exist in cases where evolutionary changes
to one trait do not disrupt the successful behavior contributed
by other traits.
For example,
robots that exhibited the locally optimal trotting behavior 
(Fig. \ref{fig:trot-gallop-roll}A)
exhibited a tight coupling between morphology and control, and thus evolution was 
unable to canalize development in either one, since mutations to one subsystem 
tended to disrupt the other.
Brief ontogenetic periods of rolling behavior 
(Fig. \ref{fig:trot-gallop-roll}C), 
on the other hand, could be temporally extended by evolution through canalization of the morphology alone
(Fig. \ref{fig:trot-gallop-roll}D), 
since these morphologies are generally robust to the pattern of actuation.
The key observation here is that only phenotypic traits that render the agent robust to changes in other traits become assimilated, a phenomenon we term differential canalization. 

This insight was exposed by modeling the development of simulated robots as they interacted with a physically realistic environment.
Differential canalization may be possible in disembodied agents as well, 
if they conform to appropriate conditions described in Supplementary Discussion.

This finding of differential canalization has important implications for the evolutionary design of artificial and embodied agents such as robots.
Computational and engineered systems generally maintain a fixed form as they behave and are evaluated.
However, these systems are also extremely brittle when confronted with slight changes in their internal structure, such as damage, 
or in their external environment such as moving onto a new terrain
\cite{french1999catastrophic,
carlson2005ugvs,
bongard2006resilient}.
Indeed, a perennial problem in robotics and AI is finding general solutions which perform well in novel environments 
\cite{koos2013transferability,
nguyen2015deep
}.
Our results demonstrate how incorporating morphological development in the optimization of robots can reveal, through differential canalization, characters which are robust to internal changes.
Robots that are robust to internal changes in their controllers may also be robust to external changes in their environment \cite{bongard2011morphological}.
Thus, allowing robots to change their structure as they behave might facilitate evolutionary improvement of their descendants, even if these robots will be deployed with static phenotypes or in relatively unchanging environments.

These results are particularly important for the nascent field of soft robotics in which engineers cannot as easily presuppose a robot's body plan and optimize controllers for it because designing such machines manually is unintuitive
\cite{lipson2014challenges, pfeifer2012challenges}.
Our approach addresses this challenge, because differential canalization provides a mechanism whereby static yet robust soft robot morphologies may be automatically discovered using evolutionary algorithms for a given task environment.
Furthermore, future soft robots could potentially alter their shape to best match the current task by selecting from previously trained and canalized forms.
This change might occur pneumatically, as in Shepherd \textit{et al.} \cite{shepherd2011multigait}, or it could modulate other material properties such as stiffness (e.g.~using a muscular hydrostat).

We have shown that 
for canalization to occur in our developmental model, some form of paedomorphosis must also occur. However, there are at least two distinct methods by which such heterochrony can proceed: progenesis and neoteny.
Progenesis 
could occur through mutations which move initial parameter values $(\ell,\, \phi)$ toward their final values $(\ell^*,\, \phi^*)$.
Neoteny 
could instead occur through mutations which move final values $(\ell^*,\, \phi^*)$ toward their initial values $(\ell,\, \phi)$.
Although a superior phenotype can materialize anywhere along the ontogenetic timeline, late onset mutations are less likely to be deleterious than early onset mutations.
This is because our developmental model is linear in terms of process, and interfering with any step affects all temporally-downstream steps. 
Since the probability of a mutation being beneficial is inversely proportional to its phenotypic magnitude \cite{fisher1930genetical}, mutational changes in the terminal stages of development require the smallest change to the developmental program.
Hence, late-onset discoveries of superior traits are more likely to occur without breaking functionality at other points in ontogeny, and these traits can become canalized by evolution through progenesis: mutations which reduce the amount of ontogenetic time prior to realizing the superior trait (by moving $\ell \rightarrow \ell^*$ and/or $\phi \rightarrow \phi^*$). 
Indeed progenesis was observed most often in our trials (Fig. \ref{fig-discovery}): late onset mutations which transform a walking robot into a rolling one are discovered by the evolutionary process, and are then moved back toward the birth of the robots'
descendants through subsequent mutations.

Finally, we would like to note the observed phenomenon of \textit{increased} 
plasticity prior to genetic assimilation.
Models of the Baldwin effect usually assume that phenotypic plasticity itself does not evolve, although it has been shown how major changes in the environment can select for increased plasticity in a character that is initially canalized \cite{lande2009adaptation}.
In our experiments however, there is no environmental change.
There is also a related concept known as `sensitive periods' of development in which an organism's phenotype is more responsive to experience 
\cite{bateson1979sensitive
}.
Despite great interest in sensitive periods, the adaptive reasons why they have evolved are unclear \cite{Fawcett2015}.
In our model, increasing the amount of morphological development increases the chance of capturing an advantageous static phenotype, which can then be canalized, once found.
However, a phenotype will not realize the globally optimal solution by simply maximizing development.
This would merely lengthen the \textit{line} on which development unfolds in phenotypic hyperspace ($n$-dimensional real space).

The developmental model described herein is intentionally minimalistic in order to isolate the effect of morphological and neurological change in the evolutionary search for embodied agents.
The simplifying assumptions necessary to do so make it difficult to assess the biological implications.
For example, we model development as an open loop process 
and thus ignore environmental queues and sensory feedback 
\cite{
Moczekrspb20110971,
snell2013overview
}.
We also disregard the costs and constraints of phenotypic plasticity 
\cite{
snell2012selective,
murren2015constraints
}. 
By removing these confounding factors, we hope these results will help generate novel hypotheses about morphological development, heterochrony, modularity and evolvability in biological systems.

\externaldocument{results}

\section*{Methods}
\label{sec:methods}

\subsection*{Ballistic development.}

Ballistic development 
$\mathcal{B}(t)$
is simply a linear function from a starting value $a$ to a final value $b$, in ontogenetic time $t\in(0,\tau)$, thus: 
\begin{equation}
\label{eq:ballistic-devo}
\mathcal{B}(t) = a + \frac{t(b-a)}{\tau}.
\end{equation}
For Evo robots, $a=b$, hence:
\begin{equation}
\label{eq:no-devo}
\mathcal{B}(t) = a,
\end{equation}
which is just a constant value in ontogenetic time.

Because $a$ and $b$ are constants set by the genotype and $\tau=10$ (sec) is a fixed hyperparameter, development is predetermined, monotonic and irreversible|in a word: ballistic.

\subsection*{Current length.} 

For smaller voxels, it is necessary to implement damping into their actuation to avoid simulation instability. 
Actuation amplitude is limited by a linear damping factor $\xi(x) = \min\{1,\, (4x-1)/3\}$, which only affects voxels with resting length less than one centimeter, and approaches zero (no actuation) as the resting length goes to its lower bound of 0.25 cm.

Actuation of the $k$-th voxel $\psi_k(t)$ therefore depends on the starting and final phase offsets $(\phi_k,\, \phi_k^*)$ for relative displacement, and on the starting and final resting lengths $(\ell_k,\, \ell_k^*)$ for amplitude. With maximum amplitude $A=0.14$ cm and a fixed frequency $f=4$ Hz, we have:
\begin{equation}
\label{eq-actuation}
\psi_k(t) = A \cdot \sin\left[2\pi f t + \phi_k + \frac{t(\phi_k^*-\phi_k)}{\tau}\right] \cdot \xi\left[\ell_k + \frac{t(\ell_k^*-\ell_k)}{\tau}\right]
\end{equation}
The current length of the $k$-th voxel at time $t$, denoted by $\mathcal{L}_k(t)$, is the resting length plus the offset and damped signal $\psi_k(t)$.
\begin{equation}
\label{eq-curr-length}
\mathcal{L}_k(t) = \ell_k + \frac{t(\ell_k^*-\ell_k)}{\tau} + \psi_k(t)
\end{equation}
Current length is broken down into its constituent parts for a single 
voxel, under each treatment, in Supplementary Fig. \hyperref[fig:S2]{S2}.

\subsection*{Evolutionary algorithm.}

We employed a standard evolutionary algorithm, Age-Fitness-Pareto Optimization \cite{Schmidt2011}, which uses the concept of Pareto dominance and an objective of age (in addition to fitness) intended to promote diversity among candidate designs. 
\textit{A single evolutionary trial maintains a population of thirty robots, for ten thousand generations.}
Every generation, the population is first doubled by creating modified copies of each individual in the population.
The age of each individual is then incremented by one.
Next, an additional random individual (with age zero) is injected into the population (which now consists of 61 robots). 
Finally, selection reduces the population down to its original size (30 robots) according to the two objectives of fitness (maximized) and age (minimized).

We performed the above procedure thirty times to produce \textit{thirty independent evolutionary trials.}
That the number of trials is the same as the population size within each trial is an admittedly confusing coincidence.

The mutation rate is also evolved for each voxel, independently, and slightly modified every time a genotype is copied from parent to child.
These 48 independent mutation rates are initialized such that only a single voxel is mutated on average.
Mutations follow a normal distribution $(\sigma_{\ell}=0.75$ cm; $\sigma_{\phi}=\pi/2)$ and are applied by first selecting what parameter types to mutate $(\phi_k,\, \phi_k^*,\, \ell_k,\, \ell_k^*)$, and then choosing, for each parameter separately, which voxels to mutate. 
In Supplemental Materials we provide exact derivations of the expected genotypic impact of mutations, in terms of the proportions of voxels and parameters modified, for a given fixed mutation rate $\lambda$.
There is a negligible difference between Evo and Evo-Devo in terms of the expected number of parent voxels modified during mutation (Supplementary Fig. \hyperref[fig:S4]{S4}).

\subsection*{Developmental windows.}

The amount of development in a particular voxel can range from zero (in the case that starting and final values are equal) to 1.5 cm for the morphology (which ranges from 0.25 cm to 1.75 cm) and $\pi$ for the controller (which ranges from $-\pi/2$ to $\pi/2$). 
The morphological development window, $W_L$, is the sum of the absolute difference of starting and final resting lengths across the robot's 48 voxels, divided by the total amount of possible morphological development.
\begin{equation}
W_L = \frac{1}{48(1.5)} \sum_{k=1}^{48} |\ell^*_k-\ell_k|
\label{eq-WL}
\end{equation}
The controller development window, $W_{\Phi}$, is the sum of the absolute difference of starting and final phase offsets across the robot's 48 voxels, divided by the total amount of possible controller development. 
\begin{equation}
W_{\Phi} = \frac{1}{48\pi} \sum_{k=1}^{48} |\phi^*_k-\phi_k|
\label{eq-WPhi}
\end{equation}

\subsection*{Statistical hypothesis testing.}

We used the Mann-Whitney $U$ test to calculate the $p$-values reported in this paper.

\subsection*{Data availability.}

\href{https://github.com/skriegman/how-devo-can-guide-evo}{\textbf{github.com/skriegman/how-devo-can-guide-evo}} contains the source code necessary for reproducing the results reported in this paper.





\bibliography{main}

\section*{Acknowledgments}


This work was supported by 
Army Research Office award W911NF-16-1-10304
and DARPA contract HR0011-18-2-0022. 
The computational resources provided by the 
Vermont Advanced Computing Core 
are gratefully acknowledged.

\section*{Author contributions statement}

S.K. conducted the experiments and wrote the manuscript.
All authors conceived the experiments, analyzed the results and reviewed the manuscript.

\section*{Competing interests}
The authors declare no competing interests.

\clearpage

\section*{Supplementary Video}
\subsection*{Supplementary Video S1.}
\label{video}
\href{https://youtu.be/Ee2sU-AZWC4}{\textbf{youtu.be/Ee2sU-AZWC4}}
provides a high-level overview of the results reported in this paper.

\section*{Supplementary Discussion}
\label{sec:supplementary-discussion}

\subsection*{Embodiment.}

We consider an agent to be embodied if its output affects its input.
This relationship may be represented by the simple update rule $\ell_{t+1} = f(\ell_t,\, \phi)$,
where $\ell_t$ denotes the morphology of an agent at time $t$, and $\phi$ denotes its control policy.
In a disembodied system, changes to the morphology are not directly constrained by its current state; the update rule becomes: $\ell_{t+1} = f(\phi)$.

Once a round robot begins rolling, its control policy cannot instantaneously force the system to go in the other direction, since
momentum will tend to preserve forward movement.
This has the effect of reducing selection pressure on the controller, since fewer variations are deleterious.
This allows evolution to continue climbing fitness gradients by mutating the controllers within these permissive body plans.

This might also be possible in disembodied agents if other dimensions of the system can be changed by some search process such as to facilitate the search for $\phi$.

\section*{Supplementary Methods}
\label{sec:supplementary-methods}

\subsection*{Rigid-bodied robots.}

Rigid-bodied robots and their environment were simulated using Pyrosim
(\href{https://ccappelle.github.io/pyrosim/}{ccappelle.github.io/pyrosim}).
The robot is a quadruped with a large, spherical abdomen; each leg is attached by a single degree-of-freedom hinge joint.\\[0.75em]
\centerline{\includegraphics[width=0.4\linewidth]{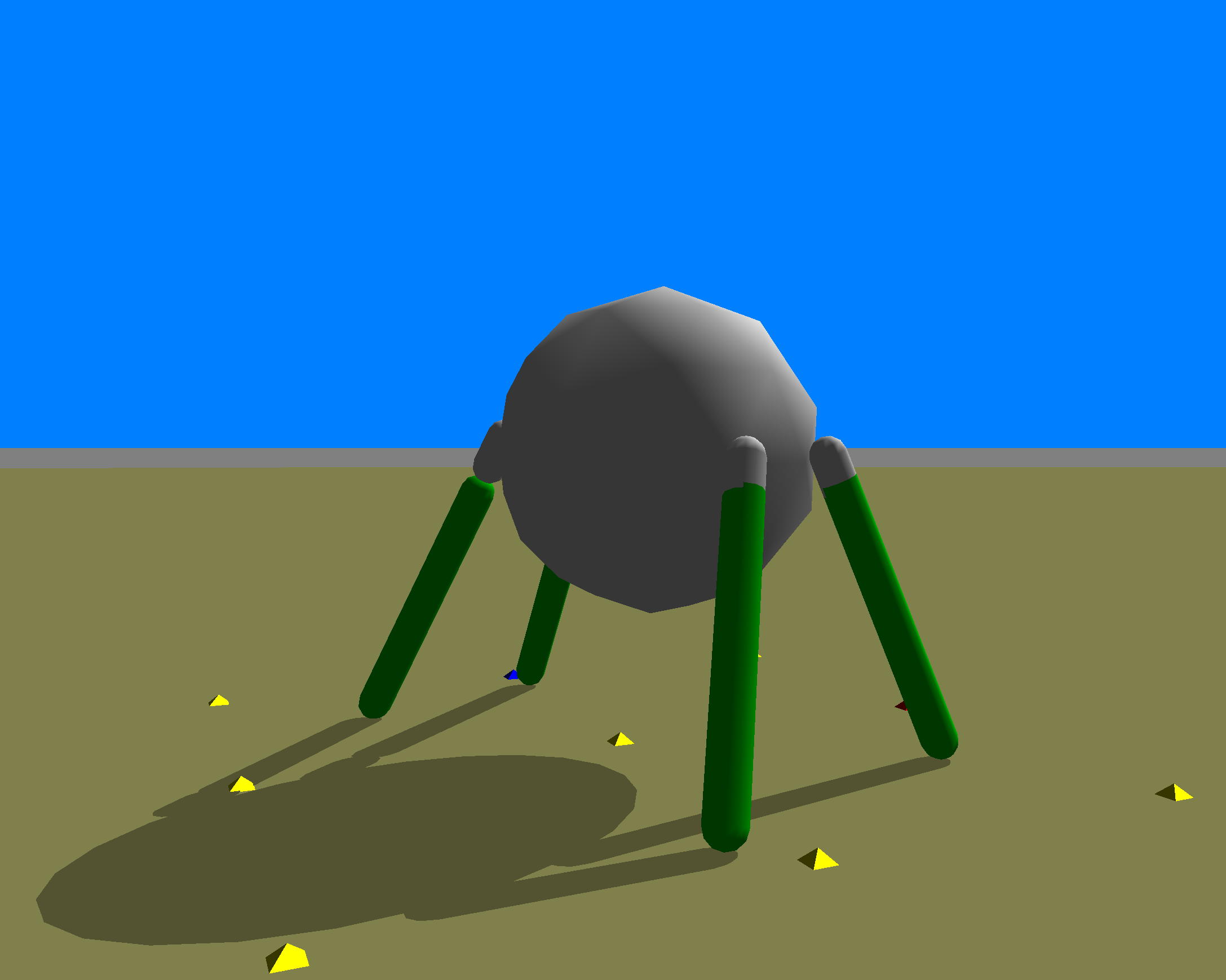}}

\vspace{0.75em}
Morphological development was approximated in rigid bodies using linear actuators to slowly lengthen or shorten the length of each leg, from an evolved starting value (between 0 and 1) to an evolved final value (between 0 and 1).
The controller is a simple neural net: two central pattern generators are fully connected to four motor neurons, each of which innervate a separate hinge joint.
Controller development was approximated in neural networks through ballistic change to each synaptic weight: As the simulation proceeds, each weight develops linearly, from an evolved starting value (between -1 and +1) to an evolved final value (between -1 and +1).

The genotype spans two arrays: one for initial and final synaptic weights (controller), and another for initial and final leg lengths (morphology). 
Mutations affect, on average, a single element in each array.
Apart from the genotype and its mutations, the evolutionary algorithm is identical to that of the soft robots.
However, the task environment now consists of a sloped floor, declined toward a light source; and
performance is measured by the average light intensity recorded by a light sensor embedded in the center of the agent's abdomen, according to the inverse square law of light propagation, at each time step in its life. Occlusion of the light caused by interference of the robot’s own body parts was not simulated.

The results are presented below in Supplementary Fig. \hyperref[fig:S5]{S5}.

\subsection*{Mutations for soft robots.}

\begin{figure}
\centering
\includegraphics[width=0.5\linewidth]{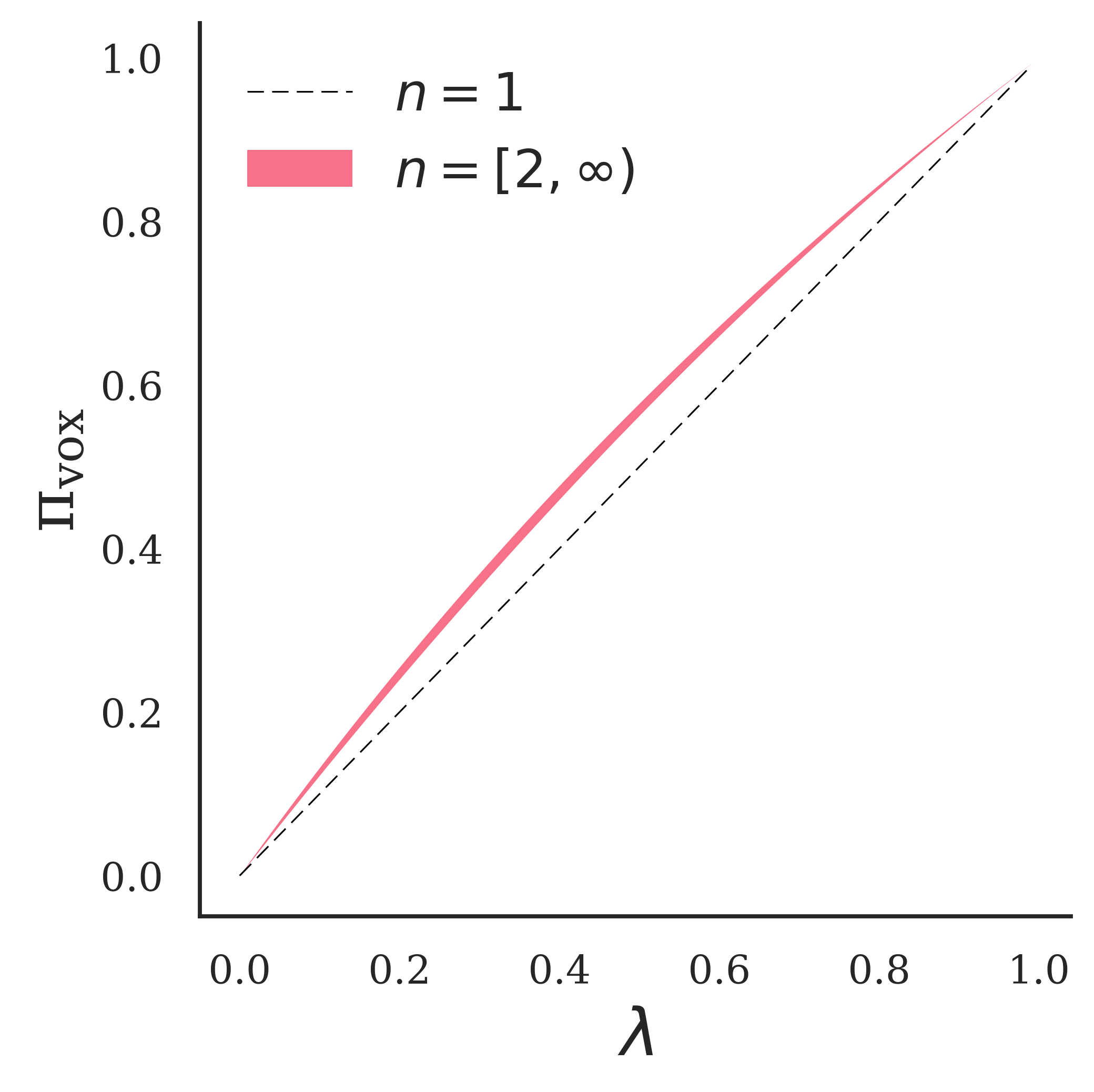}
\caption*{\label{fig:S4}\textbf{{\sf \textbf{Figure S4.}} Mutational impact.} The expected proportion of voxels modified, $\pi_{\text{vox}}$, where $n$ is the number of material properties that can be mutated, and $\lambda$ is the mutation rate.
A derivation is provided in Supplementary Methods.
Regardless of $n$,
when $\lambda=1$, every voxel must be mutated, and when $\lambda=0$, no voxels can be mutated. 
Between these two points, there is an extremely tight bound on the proportion of voxels mutated for any $n>1$. 
In this paper, we have treatments Evo, with $n=2$, and Evo-Devo, with $n=4$.
}
\end{figure}

The following derivation shows that there is a negligible difference in the mutations produced by the Evo and Evo-Devo treatments, in terms of the number of voxels modified (Fig. \hyperref[fig:S4]{S4}). 

Each voxel cell of a soft robot has its own material properties that can be changed by the evolutionary algorithm.
Evo voxels have two material properties: (1) resting length and (2) phase offset.
Evo-Devo voxels have four material properties: (1) initial resting length, (2) final resting length, (3) initial phase offset, and (4) final phase offset.

Mutations are applied by first choosing which material properties to mutate, and then choosing, separately for each property, which voxels to modify.
For each of the $n$ material properties, we select it with independent probability $p=1/n$. 
If none are selected, we randomly choose one. This occurs with probability $\left(1-p\right)^n$.
Hence the number of selected material properties for mutation is a random variable $S$ which follows a truncated binomial distribution,
\begin{equation}
\text{Pr}(S = s \mid n) = 
	\begin{cases} 
        0 & \text{ for } \;  s = 0 \\[4pt]
        n p(1-p)^{n-1} + (1-p)^n & \text{ for } \;  s = 1 \\[8pt]
        \dbinom{n}{s} p^s (1-p)^{n-s} & \text{ for } \;  s > 1
	\end{cases}
\end{equation}
The expected number of selected material properties is then:
\begin{align}
\mathbb{E}(S) 
&=
np(1-p)^{n-1} + (1-p)^n + \sum_{s=2}^n s \binom{n}{s} p^s (1-p)^{n-s} \\
&= 
(1-p)^n + \sum_{s=1}^n s \binom{n}{s} p^s (1-p)^{n-s} \\
&= 
(1-p)^n + np \\
&= 
(1-p)^n+1 .
\end{align}

For a selected material property, 
each voxel is mutated independently with probability $\lambda$, a hyperparameter we call the \textbf{mutation rate}.
The expected \textit{number} of genotype elements mutated given $K$ total voxels is thus:
\begin{equation}
\delta_{\text{gene}} = \lambda K \cdot \mathbb{E}(S) .
\end{equation}
Dividing by the length of the genome, $nK$, we get the expected \textit{proportion} of genotype elements mutated:
\begin{equation}
\pi_{\text{gene}} = \lambda/n \cdot \mathbb{E}(S).
\end{equation}
Note that imposing bilateral symmetry does not change these expected values.

We have $K=48$ total voxels, and $n=\{2,4\}$ material properties for our two main experimental treatments \{Evo, Evo-Devo\}, respectively. The expected difference between a robot and its offspring, in terms of genotype elements, is summarized in the following table.
\vspace{-1em}
\begin{table}[!ht]
\centering
\begin{tabular}{rcc}
 & $n=2$ & $n=4$  \\ \cline{2-3} 
\multicolumn{1}{r|}{$\delta_{\text{gene}}$}     & \multicolumn{1}{c|}{$60\lambda$}    & \multicolumn{1}{c|}{$63.8175\lambda$} \\ \cline{2-3} 
\multicolumn{1}{r|}{$\pi_{\text{gene}}$} & \multicolumn{1}{c|}{$0.625\lambda$} & \multicolumn{1}{c|}{$0.3291\lambda$}  \\ \cline{2-3} 
\end{tabular}
\end{table}

\noindent
However, because multiple material properties can be mutated within a single voxel, the expected number of voxels mutated is lower than the expected number of genotype elements mutated. 
To calculate the average number of voxels mutated we need to consider a hierarchy of binomial distributions.

Given that $S$ material properties were selected for mutation,
the number of material properties mutated within a single voxel, $M$ follows a binomial distribution, 
\begin{equation}
\text{Pr}(M=m \mid S, \lambda) 
= 
\dbinom{S}{m} \lambda^m (1-\lambda)^{S-m} .
\end{equation}
For brevity, let's denote the probability that at least one mutation occurs within the voxel as $\theta$,
\begin{equation}
\theta = \text{Pr}(M>0 \mid S,\lambda)
=
1-(1-\lambda)^S .
\end{equation}
Then the number of voxels mutated, $V$, across a total of $K$ voxels and $S$ selected material properties, also follows a binomial distribution:
\begin{equation}
\text{Pr}(V=v \mid S, K, \lambda, n) 
= 
\dbinom{K}{v} \theta^v (1-\theta)^{K-v}.
\end{equation}
And the expected number of voxels mutated (out of $K$ total) is
\begin{align}
\delta_{\text{vox}}
&=
\mathbb{E}(V \mid K, \lambda, n) \\
&=
\mathbb{E}_S \, \mathbb{E}_{V}(V \mid S, K, \lambda, n) \\
&= 
\mathbb{E}_S (K\theta \mid S, K, \lambda, n) \\
&=
K \left\{
1-\mathbb{E}_S \left[(1-\lambda)^S \mid \lambda, n\right] \right\} \\
&=
K \left\{ 1 - \left[ (1-\lambda)(1-p)^n
+  \sum_{s=1}^{n} (1-\lambda)^s \dbinom{n}{s} p^s (1-p)^{n-s}
 \right] \right\} \\
&=
K \left\{ 
1 - (1-p)^n \left[ \left(\frac{\lambda p-1}{p - 1}\right)^n - \lambda\right]
\right\} 
\end{align}

There is an extremely tight bound on the proportion of voxels mutated, $\pi_{\text{vox}} = \delta_{\text{vox}}/K$, for any $n > 1$  (Fig. \hyperref[fig:S4]{S4}).
Thus mutations in Evo $(n=2)$ and Evo-Devo $(n=4)$ have practically the same impact in terms of the number of voxels modified.
For completeness, the following table displays $\delta_{\text{vox}}$ for the specific values of $\lambda$ considered by our hyperparameter sweep ($K=48$) (Fig. \hyperref[fig:S3]{S3}). 
\vspace{-0.5em}
\begin{table}[h!]
\centering
\begin{tabular}{cccccccccc}
  & & \multicolumn{8}{c}{$\lambda$} \\ \cline{3-10}
  & \multicolumn{1}{c|}{}  & 1/48    & 2/48    & 4/48    & 8/48    & 16/48   & 24/48   & 32/48   & \multicolumn{1}{c|}{48/48} \\ \cline{2-10} 
\multicolumn{1}{c|}{\multirow{2}{*}{$n$}} & \multicolumn{1}{c|}{2} &
1.25 & 2.48 & 4.92 & 9.67 & 18.67 & 27 & 34.67 & \multicolumn{1}{c|}{48}    \\
\multicolumn{1}{c|}{}                     & \multicolumn{1}{c|}{4} & 
1.3 & 2.6 & 5.14 & 10.05 & 19.17 & 27.46 & 34.98 & \multicolumn{1}{c|}{48}    \\ \cline{2-10} 
\end{tabular}
\end{table}

\section*{Supplementary References}
\label{supp-refs}
1. Ancel, L. W. Undermining the baldwin expediting effect: does phenotypic plasticity accelerate evolution? \textit{Theor. population
biology} \textbf{58}, 307–319 (2000).

\externaldocument{methods}

\begin{figure}
\centering
\includegraphics[width=\linewidth]{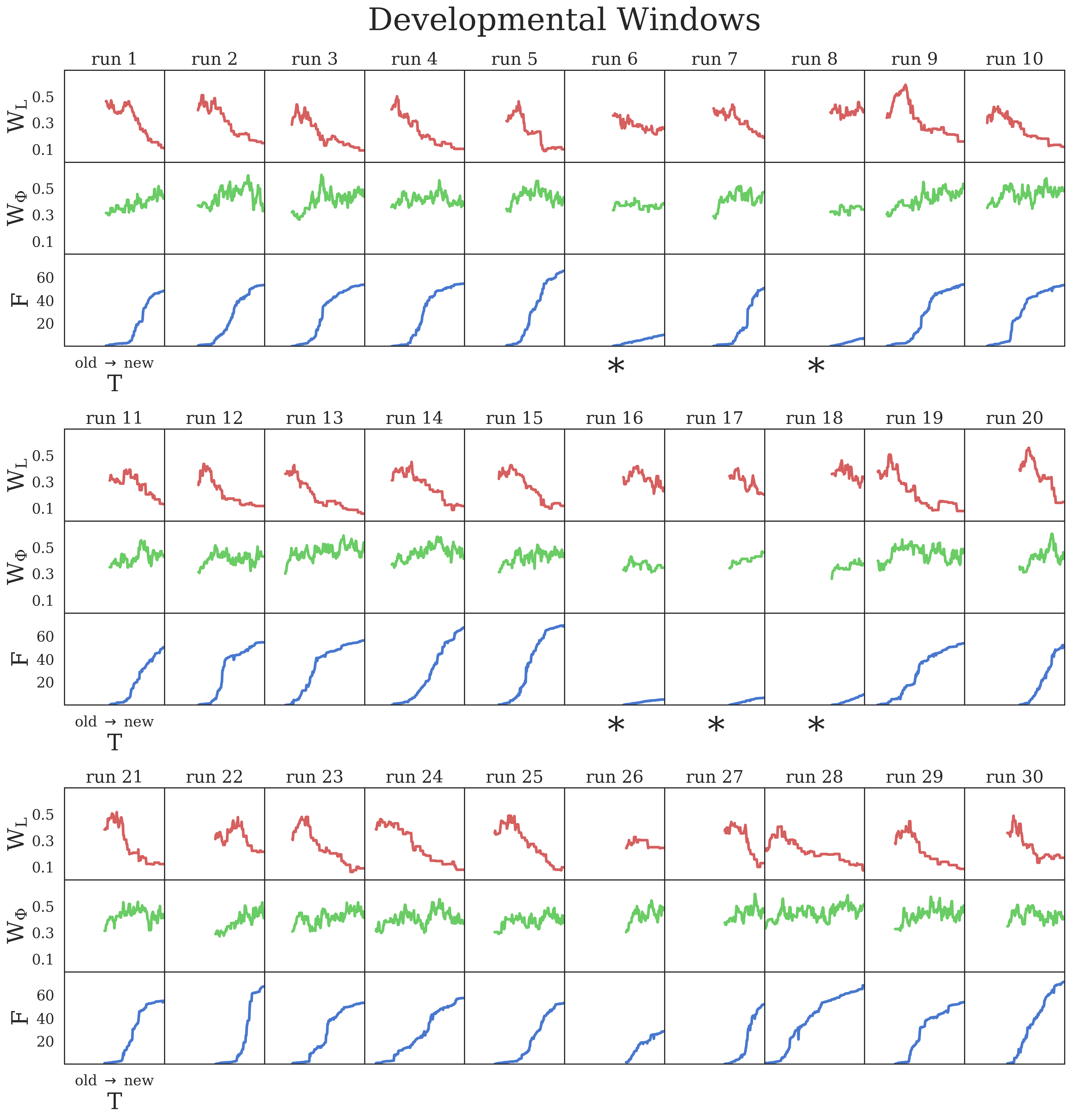}
\caption*{\label{fig:S1}\textbf{{\sf \textbf{Figure S1.}} Evolutionary change during 30 Evo-Devo trials.} The amount of morphological development, $W_L$ (see Equation 3), controller development, $W_{\Phi}$ (see Equation 4), and fitness, $F$, for the lineages of the 30 Evo-Devo run champions. Evolutionary time, $T$, moves from the oldest ancestor (left) to the run champion (right). A general trend emerges wherein lineages initially increase their morphological development in $T$ (rising red curves) and subsequently decrease morphological development to almost zero (falling red curves). Five of the 30 evolutionary trials, annotated by {\Large $\ast$}, fell into a local optima.}
\end{figure}

\begin{figure}
\centering
\includegraphics[width=0.64\linewidth]{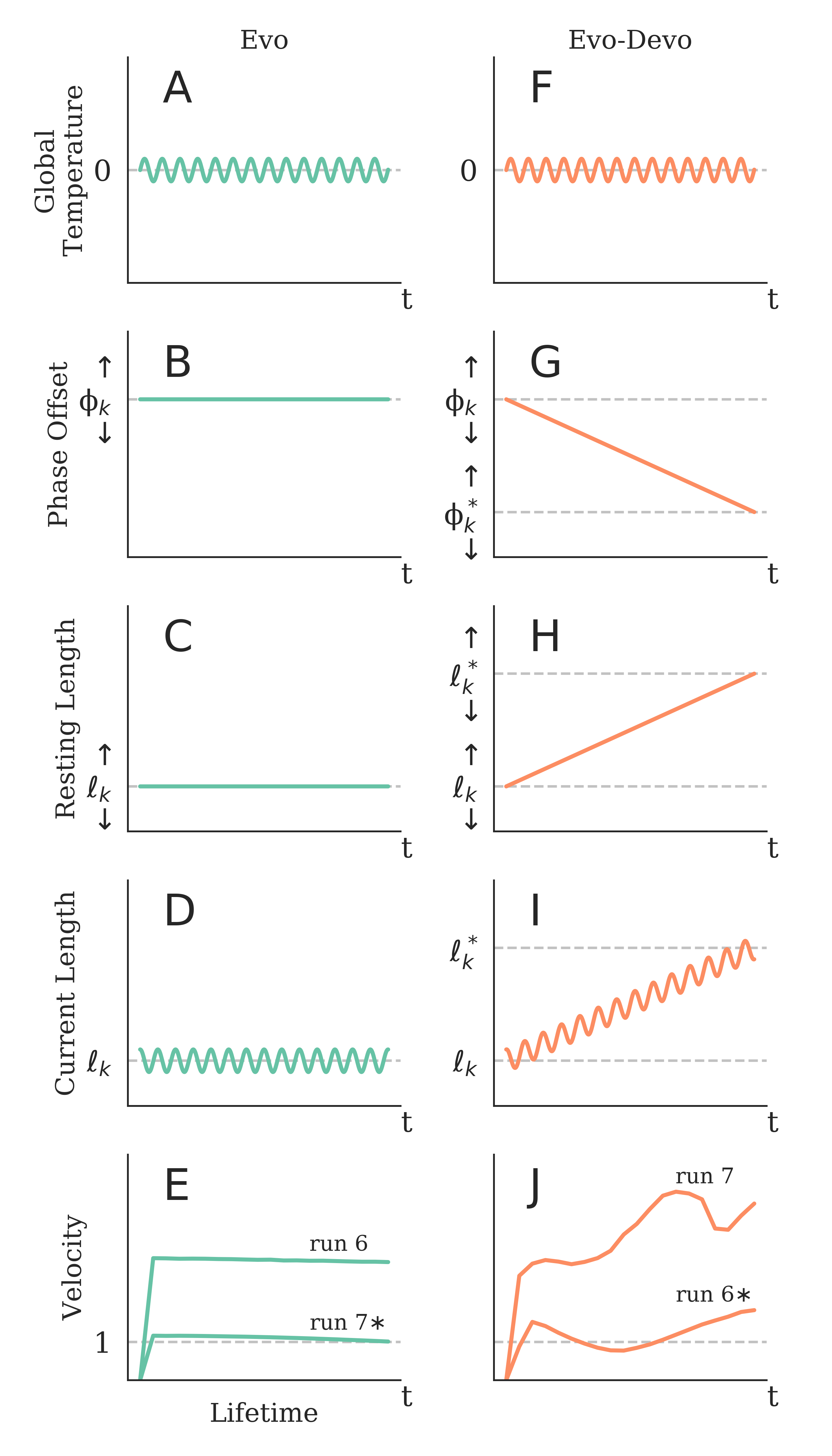}
\caption*{\label{fig:S2}\textbf{{\sf \textbf{Figure S2.}}  Experimental treatments.} The phase of an oscillating global temperature (A, F) is offset in the $k$-th voxel by a linear function from $\phi_k$ to $\phi_k^*$ (B, G). 
The resting length of the $k$-th voxel is a linear function from $\ell_k$ to $\ell_k^*$ (C, H). 
For Evo, there is no development, so $\phi_k=\phi_k^*$ and $\ell_k=\ell_k^*$.
The offset actuation is added on top of the resting length to give the current length of the $k$-th voxel (D, I). 
These example voxel-level changes occur across ontogenetic time $(t)$, independently in each of the 48 voxels, and together interact with the environment to generate robot-level velocity (E, J).
To see this, we averaged displacement across intervals of two actuation cycles (0.5 sec) and plotted the smoothed velocities for two Evo-Devo run champions with minimal canalization (J) alongside their non-developmental counterparts (E).
Also note that the frequency of actuation is plotted here at 1.4 Hz; but, in our experiments, we used a frequency of 4 Hz.
}
\end{figure}

\begin{figure}
\centering
\includegraphics[width=0.65\linewidth]{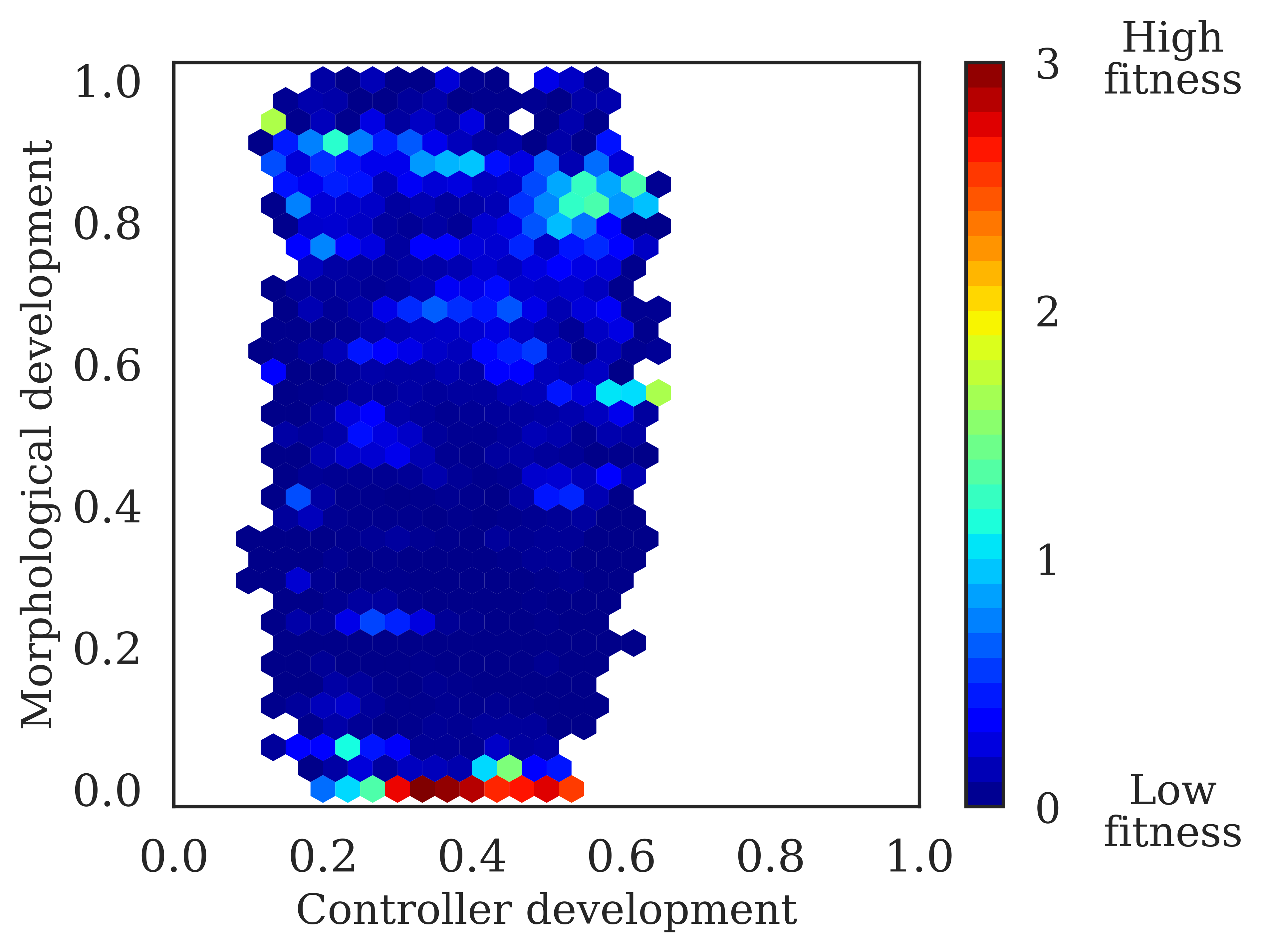}
\caption*{\label{fig:S5}\textbf{{\sf \textbf{Figure S5.}} Differential canalization in rigid bodies.} 
This environment consists of a sloped floor, declined toward a light source. 
Rigid-bodied robots, which are described in Supplementary Methods, perform phototaxis.
Although longer legs produce faster walking behaviors, the shortest leg setting places the robot's large, spherical abdomen onto the ground, allowing the robot to simply roll down the ramp toward the light. 
An ancestor discovers rolling over toward the end of its evaluation period through ontogenetic morphological change that compresses leg lengths. 
This rollable morphology is assimilated to the start of development through differential canalization: The sooner a robot shrinks its legs, the sooner it can begin rolling; eventually descendants start with compressed legs, are able to immediately roll, exhibit little to no morphological change, but continue to sweep over many different synaptic weights as they behave. 
Rolling down the slope is not entirely passive, however, since protruding legs, even at their shortest setting, can slow down or even stop forward movement. 
The best controllers not only avoid such interference, but also guide rolling toward the light source. 
However, development in this particular case did not affect evolvability. 
These results are consistent with predictions made by other quantitative models used in theoretical biology that suggest that plasticity only expedites evolution under restrictive conditions\hyperref[supp-refs]{$^1$}. 
We hypothesize that, in our case, this is because the search space is not deceptive enough: Once the rigid-bodied robot evolves to compress its legs, and touch its abdomen to the sloped floor, it will tend to roll for the remainder of its evaluation period.
That is, in contrast with the soft robots, there is no intermediate stage between walking and rolling that suffers the fitness penalty of no longer being able to move.
}
\end{figure}

\end{document}